\documentstyle[12pt]{article}
\def\rank{\mathop{\rm rank}\nolimits}
\def\pa{\partial}

\def\nta{\not\!\!{\tilde A}}

\makeatletter
\@addtoreset{equation}{section}
\makeatother
\title{Weak Dirac bracket construction and the superparticle\\
covariant quantization problem}
\author{A.A.Deriglazov\thanks{E-mail: deriglaz@phys.tsu.tomsk.su},
A.V.Galajinsky, S.L.Lyakhovich}
\date{Department of Theoretical Physics,\\
Tomsk State University\\
Tomsk 634050, Russia}
\begin{document}
\maketitle
\begin{abstract}
The general procedure of constructing a consistent covariant Dirac-type
bracket for models with mixed first and second class constraints is
presented. The proposed scheme essentially relies upon explicit
separation of the initial constraints into infinitely reducible
first and second class ones (by making use of some appropriately constructed
covariant projectors). Reducibility of the second class constraints
involved manifests itself in weakening some properties of the bracket
as compared to the standard Dirac one. In particular, a commutation of
any quantity with the second class constraints and the Jacobi identity
take place on the second class constraints surface only. The developed
procedure is realized for $N=1$ Brink--Schwarz superparticle in arbitrary
dimension and for $N=1,D=9$ massive superparticle with Wess--Zumino term.
A possibility to apply the bracket for quantizing the superparticles
within the framework of the recent unified algebra approach by Batalin
and Tyutin [20--22] is examined. In particular, it is shown that for
$D=9$ massive superparticle it is impossible to construct Dirac-type
bracket possessing (strong) Jacobi identity in a full phase space.
\end{abstract}
\section{Introduction}
The superparticle covariant quantization problem has long been realized
[1--4] to consist in adequate extension of the initial phase space [5].
There were a number of attempts in this direction. The most successful
approaches to date are twistor-like formulations [3, 6--10], harmonic
superspace technique [11--15] and the null-vectors approach of Ref. 2.
Another sight to the problem lies in the fact that, instead of
quantizing the original Brink--Schwarz theory, it is constructed the
superparticle model [16, 17 and references therein] which will lead
after quantization to the covariant SYM.

The key idea of the harmonic superspace approach [13] was to introduce
additional harmonic variables, which played the role of a bridge between
the $SO(1,D-1)$ indices and some internal space indices, to split
the initial fermionic constraints into the first and second class parts.
Then twistor-like variables might be used [3, 13] to convert the second
class constraints into the first class ones, what brought the theory to
the form admitting conventional canonical quantization.\footnote{We mostly
discuss $N=1,D=10$ case for which manifestly covariant quantization is the
principal problem.} However, the introduced auxiliary variables
turn out to be various for different dimensions [3, 12]. The resulting
constraint system can be irreducible or infinitely reducible depending
on the dimension [3]. In the latter case there arises an additional serious
problem being connected with constructing a functional integral for
the model [2, 3]. Within the framework of operator quantization, the
wave functions depend on twistor or harmonic variables what makes
understanding the results in terms of ordinary (super)fields difficult
(in the special case of a compact Lorentz-harmonic superspace, however,
the problem can be solved [12]). Additional source of difficulties
lies in the general status of the conversion method itself.
Actually, although the approach is known for a long time [27], it still
remains unclear whether a system after conversion is always
physically equivalent to the original one. The general formalism may
now offer a proof of the equivalence which is essentially local [28].
Such a consideration is enough for the case of a conventional
perturbative field theory, but it is likely unable to take into account the
effect of the reduced phase space global geometry which may have a
significant influence on a physical spectrum of the particle model. In
view of the all mentioned problematic points of the conversion
method it seems interesting to study an alternative approach dealing with
the superparticle in its original formulation [5]. In this connection,
it is relevant to mention the ``unified algebra'' approach recently
developed [20--22] just for the systems with mixed first- and
second-class constraints. In principle, the proposed construction does
not require an explicit separation of the first and second class constraints
(what is just the basic problem of the superparticle, superstring models).
An application of the procedure for concrete theories, however, implies
the existence of some classical bracket (with {\it all} the rank and
algebraic properties of the standard Dirac one) as a boundary condition
to the basic generating equations [21]. Although the general
construction does not include an algorithm of building this
bracket, it is implied to be known ``from the outset''.

In the present paper we propose the general scheme of constructing a
consistent covariant Dirac-type bracket for models with mixed first
and second class constraints. A possibility to apply this
bracket in the context of the quantization method developed in
Refs. 20--22 is examined in the work.

There are two natural ways of building the bracket with needed
properties. First of them consists in splitting the initial constraints
into infinitely reducible first and second class parts (by  making  use
of some covariant projectors) and subsequent generalizing the standard
Dirac bracket construction to the case of infinitely reducible second
class constraints. The second line is to write down the most general
ansatz for the bracket and then to require all needed rank and
algebraic properties for the construction (what will specify the
coefficient functions of the ansatz).

Possibilities to construct the brackets of the first kind for the
superparticle, superstring models were examined in Refs. 2, 11, 18, 19,
and 25. It seems surprising but the Jacobi identity is problematic for
each of the suggested brackets. Actually, within the framework of
the adopted scheme [2, 4] the only property\footnote{Correct
elimination of the redundant second class constraints implies as well
linear independence of the reducible first and second class
constraints in the question (see Ref. 4).} to be satisfied by the
bracket was a weak commutation of any quantity with the second class
constraints involved, i.e., only rank conditions were taken into
account. Among the algebraic properties, the graded symmetry, linearity
and differentiation took place by the construction while the
Jacobi identity was not embedded into the scheme in some special way. In
the general case, a bracket of such a kind may possess the identity in
a strong sense (off the constraint surface), weakly (on the constraint
shell) or has no the property at all. For instance, the Jacobi identity
for the bracket of Ref. 19 takes place on the second class constraint
surface only (see subsec. 3.2) and it fails to be fulfilled off-shell.
As will be shown (Secs. 2 and 4) the weakening the identity is an
essential ingredient of the consistent covariant Dirac-type bracket
associated with the infinitely reducible second class constraints.

In this context, the question of the bracket with the strong Jacobi
identity arises naturally. It turns out that it is the second approach
to the building the bracket which allows to investigate the question
completely. As an example, we consider $N=1,D=9$ massive superparticle
with Wess--Zumino term [19]. In this case it proves to be possible to
write down the most general ansatz for the fundamental phase variable
brackets provided with the correct rank conditions (Sec. 4). Requirement
of the strong Jacobi identity for the brackets implies some
restrictions on the coefficient functions of the ansatz. Our result
here is partly surprising. As will be show, there arises a
contradictory system of equations for the coefficients if the latter
were taken in the Poincar\'e covariant form, i.e. a Poincar\'e
covariant closing the identity is impossible. Thus the weakening some
properties of the bracket seems to be an essential ingredient of the
covariant description when dealing with the models possessing mixed
first and second class constraints.

The paper is organized as follows. In Sec. 2 the general procedure of
building a covariant Dirac-type bracket for the models with mixed
first and second class constraints is presented. Constructing
such a bracket turns out to be equivalent to solving certain system of
matrix equations in the enlarged phase space. As will be shown
reducibility of the second class constraints (which is a price paid for
the explicit covariance) manifests itself in weakening some properties of
the bracket as compared to the standard Dirac one. Namely, a
commutation of any quantity with the second class constraints and the
Jacobi identity take place on the second class constraints surface only
(note that this weakening still is compatible with the Dirac's
prescriptions of quantization).

In Sec. 3 the proposed construction is realized for $N=1$ Brink--Schwarz
superparticle in arbitrary dimension, and for $N=1, D=9$ massive
superparticle with Wess--Zumino term. In the first case the original
phase space must be enlarged to include one vector variable (and its
conjugate momentum) only. In the second case the bracket can be
formulated in the initial phase space and our general construction
reproduces here the results of Ref. 19. It is interesting to note
that within the framework of the developed scheme the manifestly covariant
(redundant) gauge fixing present no a special problem.

In Sec. 4 the constructed brackets are examined in the context of the
quantization procedure of Refs. 20--22. A possibility to continue the
Jacobi identity off the constraint surface (what is an essential
ingredient of the classical counterpart of the quantum bracket in Ref. 21)
is considered. As will be shown for the particular example of $D=9$ massive
superparticle, a Poincar\'e covariant extension of the bracket up to one
with the strong Jacobi identity is impossible. Thus,
although the classical boundary conditions for the unified constraint
dynamics being applied to the superparticle model can be constructed,
covariant quantum realization of the quantities turns out to be
problematic. In the Conclusion we summarize our results.

\section{General construction}

The essential ingredient of the both superparticle [5] and superstring
[23] theories is Siegel local fermionic symmetry [24], which eliminates
unphysical degrees of freedom and provides absence of negative norm
states in the quantum spectrum of the models [2]. The constraint system
of the theories in the Hamiltonian formalism includes some set of
bosonic first class constraints, which we denote as $T_A\approx0$ ($A$
is a condensed index) as well as fermionic constraints $\chi_\alpha\approx0$,
$\alpha=1,\dots,n$, from which half is the first class (the generators of
the Siegel transformations) and another half is second class, i.e.,
$\{\chi_\alpha,\chi_\beta\}\equiv\Delta^*_{\alpha\beta}$ and $\rank
\Delta^*|_{T,\chi\approx0}=n/2$. The symbol $\{A,B\}\equiv{\stackrel
\leftarrow\pa}_iA\omega^{ij}{\stackrel\to\pa}_jB$ is used to
denote the canonical Poisson bracket on a phase supermanifold, where
${\stackrel\to\pa}_i\equiv {\stackrel\to\pa}/\pa\Gamma^i$ and
$\Gamma^i$ are the local coordinates.

The main obstacle for operator covariant quantization of the models
lies in the fact that in the original phase space it is impossible
to separate the mixed first and second class constraints
$\chi_\alpha\approx0$ in a covariant irreducible way [1, 2] while the
redundant splitting may present a nontrivial task [2, 18]. To avoid this
difficulty we enlarge the initial phase space $\Gamma$ up to
$\Gamma^*\equiv (\Gamma,\Gamma_{\rm add})$, where $\Gamma_{\rm add}$ is
some set of additional variables. The new variables are implied to be
pure gauge and, consequently, there must be constraints eliminating
$\Gamma_{\rm add}$ part of $\Gamma^*$. Denote first and second class
constraints of such a kind as $\varphi_{A_1}(\Gamma^*)\approx0$ and
$\psi_{A_2}(\Gamma^*)\approx0$, $A_1=1,\dots,n_1$, $A_2=1,\dots,n_2$,
respectively. In what follows, we admit these constraints to be linearly
independent. Further, in the extended phase space we suppose the
existence of a pair of (strong) projectors $p^{\pm\beta}_\alpha(\Gamma^*)$
\begin{equation}
\begin{array}{ll} {p^+}^2=p^+, &\qquad {p^-}^2=p^-,\\
p^+p^-=0, &\qquad p^++p^-=1\end{array}
\end{equation}
splitting the original mixed constraints $\chi_\alpha$ into redundant
first and second class pieces $\chi^+\equiv p^+\chi$ and $\chi^-\equiv
p^-\chi^-$
\begin{equation}
\begin{array}{c} \{\chi^+_\alpha,\chi^+_\beta\}\approx0, \qquad
\{\chi^+_\alpha,\chi^-_\beta\}\approx0,\\
\{\chi^-_\alpha,\chi^-_\beta\}\equiv\Delta_{\alpha\beta}\approx
(p^-\Delta^*p^-)_{\alpha\beta}.\end{array}
\end{equation}
{}From Eqs. (2.2) it follows the defining equation for the projector
operators
\begin{equation}
\Delta^* p^+\approx0.
\end{equation}
Several remarks are relevant here. First, in consequence of the
identities
\begin{equation}
p^+\chi^-\equiv0, \qquad p^-\chi^+\equiv0
\end{equation}
there are only half linearly independent constraints among the conditions
$\chi^\pm_\alpha\approx0$. This means as well that the criterion of
Ref. 4 for consistent elimination of redundant second class constraints
is automatically satisfied within the framework of our construction.
Secondly, a proof of the equivalence between the separated first and
second class constraints and the initial mixed constraint system is
evident (from $\chi\approx0$ it follows $\chi^+\approx0$, $\chi^-\approx0$,
and vice versa). Thirdly, $\rank\Delta|_{T,\chi,\varphi,\psi\approx0}=n/2$
(as a consequence of Eq. (2.3)) what correctly reflects reducibility of the
resulting second class constraints. Note as well that noncovariant
projectors may always be constructed by making use of initial phase space
variables only. Actually, for the models concerned one can
find $n/2$ linearly-independent (weak) null-vectors for the matrix
$\Delta^*$: $\Delta^*_{\alpha\beta}{C^\beta}_a(\Gamma)\approx0$,
$a=1,\dots,n/2$. Then the following algebraic system of equations
${Q^a}_\gamma{C^\gamma}_b={\delta^a}_b$ can always be solved for unknown
${Q^a}_\gamma(\Gamma)$. The quantities ${p_+}^\alpha{}_\beta\equiv
{C^\alpha}_a{Q^a}_\beta$, ${p_-}^\alpha{}_\beta\equiv {\delta^\alpha}_\beta-
{C^\alpha}_a{Q^a}_\beta$ prove to be the needed noncovariant projectors.
The task of finding the covariant projectors is less trivial, some examples
will be considered in Sec. 3.

The next step of the construction is building the generalized Dirac
bracket which would allow correct elimination of the redundant second
class constraints (i.e., which is compatible with setting all the
second class constraints strongly to zero). For this aim let us find
the (symmetric) matrix $\tilde\Delta^{\alpha\beta}$ which is inverse
to $\Delta_{\alpha\beta}$ in the following sense:
\begin{equation}
\Delta\tilde\Delta=p^-,
\end{equation}
and suppose the conditions for the second class constraints
$\psi_{A_2}\approx0$:
\begin{equation}
\{\chi^-_\alpha,\psi_{A_2}\}=0, \qquad
\{\chi^+_\alpha,\psi_{A_2}\}\approx0
\end{equation}
to be satisfied.\footnote{Equations (2.6) are only technical restrictions
allowing to write down the bracket (2.7) in the simplest form. These
conditions prove to be fulfilled for all examples considered below.} On
the basis of these assumptions one can write down the following
ansatz for the Dirac-type bracket:
\begin{equation}
\{A,B\}_D=\{A,B\}-\{A,\chi^-_\alpha\}\tilde\Delta^{\alpha\beta}\{\chi^-_\beta,
B\}
-\{A,\psi_{A_2}\}\tilde\nabla^{A_2C_2}\{\psi_{C_2},B\},
\end{equation}
where $\tilde\nabla$ is the inverse matrix to $\nabla_{A_2B_2}=
\{\psi_{A_2},\psi_{B_2}\}$, $\tilde\nabla\nabla=1$. For concrete models the
brackets like Eq. (2.7) were also considered in Refs. 2, 11, 18, 19, and 25.
They differ by the choice of the constraint set $\varphi_{A_1}$,
$\psi_{A_2}$, (i.e., $\Gamma_{\rm add}$) and by the form of the
operator extracting the second class constraints. Note that we require
the operators splitting the first and second class constraints to be
strong projectors, what ensures the equivalence of the separated and
initial constraint systems. The form of Eq. (2.5) is also crucial in our
approach. Let us briefly discuss the basic properties of the bracket.
First, any quantity $A$ (weakly) commutes with the second class
constraints under the bracket (2.7)
\begin{equation}
\{A,\psi_{B_2}\}_D=0, \qquad \{A,\chi^-_\alpha\}_D=
\{A,p^-_\alpha{}^\beta\}\chi^-_\beta\approx0.
\end{equation}
On the geometrical language it means that the matrix constructed from
the fundamental brackets
\begin{equation}
p^{ij}\equiv\{\Gamma^{*i},\Gamma^{*j}\}_D=\omega^{ij}-
\omega^{ik}{\stackrel\to\pa}_k\chi^-_\alpha\tilde\Delta^{\alpha\beta}
{\stackrel\leftarrow\pa}_s\chi^-_\beta\omega^{sj}-\omega^{ik}
{\stackrel\to\pa_k}\psi_{A_2}\tilde\nabla^{A_2B_2}
{\stackrel\leftarrow\pa}_s\psi_{B_2}\omega^{sj}
\end{equation}
is degenerate (${\rm corank}\,p^{ij}=n_2+n/2$) and the weak eigenvectors
corresponding to zero eigenvalue
\begin{equation}
p^{ij}\pa_j\psi_{A_2}=0,\qquad p^{ij}\pa_j\chi^-_\alpha\approx0
\end{equation}
are normals to the second class constraints surface. It is interesting
to note that the latter relation in Eq. (2.10) can be strengthened
to yield a strict equality
\begin{equation}
p^{ij}p^-_\alpha{}^\beta\pa_j\chi^-_\beta=0.
\end{equation}
Secondly, there is a natural arbitrariness in definition of the bracket
(2.7): addition of any polynomial in the second class constraints to the
ansatz (2.7) does not change the (weak) properties of the bracket.
Thirdly, by construction the proposed bracket possesses the graded
symmetry, linearity and differentiation. The only property to be
discussed specially is the Jacobi identity. To clarify this question
consider the graded cycle
\begin{equation}
(-1)^{\epsilon_i\epsilon_k}p^{il}{\stackrel\to\pa}_lp^{jk}+
(-1)^{\epsilon_i\epsilon_j}p^{jl}{\stackrel\to\pa}_lp^{ki}+
(-1)^{\epsilon_k\epsilon_j}p^{kl}{\stackrel\to\pa}_lp^{ij}
\end{equation}
which identically vanishes in the case of the ordinary Dirac bracket and
provides the Jacobi identity for the construction. The direct proof of the
identity in that case is actually based on the fact that the matrix of
the second class constraints, say $M_{\alpha\beta}$, is invertible
$M_{\alpha\beta}\tilde M^{\beta\gamma}={\delta_\alpha}^\gamma$ and
consequently $\pa_i\tilde M=-\tilde M(\pa_iM)\tilde M$. In the case of
infinitely reducible second class constraints we deal with equation
$\Delta_{\alpha\beta}\tilde\Delta^{\beta\gamma}={p^-}_\alpha{}^\gamma$,
from which it follows the equality
\begin{equation}
{\stackrel\to\pa}_l\tilde\Delta^{\alpha\beta}=-\tilde\Delta^{\alpha
\gamma}{\stackrel\to\pa}_l\Delta_{\gamma\delta}\tilde\Delta^{\delta
\beta}+p^+_\gamma{}^\beta{\stackrel\to\pa}_l\tilde\Delta^{\gamma\alpha}
+p^+_\gamma{}^\alpha{\stackrel\to\pa}_l\tilde\Delta^{\gamma\beta}-
{\stackrel\to\pa}_l(\tilde\Delta^{\alpha\gamma}p^+_\gamma{}^\beta).\label{eq}
\end{equation}
The new terms in the right hand side of this expression are manifestation
of reducibility of the constraints involved and they actually are the source
of breaking the Jacobi identity. It can be directly verified that the
identity for the bracket (2.7) is not fulfilled. We can show,
however, that under some additional assumptions the Jacobi identity
takes place on the second class constraint surface $\chi^-_\alpha\approx0$.
The needed suppositions turn out to be of the form
\begin{equation}
\begin{array}{cc}
\{p^-_\alpha{}^\beta,\chi_\gamma\}=0,&\qquad
\{p^-_\alpha{}^\beta,p^-_\gamma{}^\delta\}=0,\\
\{\psi_{A_2},p^-_\alpha{}^\beta\}=0,&\qquad
\{\pa_i\psi_{A_2},\chi^-_\alpha\}=0,\end{array}
\end{equation}
whence it follows
\begin{equation}
\{p^+_\alpha{}^\beta,\chi_\gamma\}=0, \qquad
\{p^+_\alpha{}^\beta,p^-_\gamma{}^\delta\}=0,\qquad
\{p^+_\alpha{}^\beta,p^+_\gamma{}^\delta\}=0;
\end{equation}
\begin{equation}
\Delta= p^-\Delta^*p^-, \qquad p^+\Delta=0.
\end{equation}
Note as well that in consequence of Eq. (2.5) the conditions
\[ p^+\tilde\Delta p^-=0, \qquad p^-\tilde\Delta p^+=0 \]
are fulfilled. This means that $\tilde\Delta$ can be represented
in the form
\[ \tilde\Delta = p^-\tilde\Delta_1 p^-+p^+\tilde\Delta_2 p^+ \]
where $\tilde\Delta_1$ is some matrix consistent with Eq. (2.5) and
$\tilde\Delta_2$ is an arbitrary matrix. The contribution of the second
term into the bracket (2.7) can always be suppressed by making use of
the natural arbitrariness containing in definition of the bracket (taking
into account the identity $p^+_\delta{}^\alpha{\stackrel\to\pa}_n\chi^-_\alpha
\equiv-({\stackrel\to\pa}_np^+_\delta{}^\alpha)\chi^-_\alpha$ one concludes
that the contribution is quadratic in the second class constraints $\chi^-$).
By this reason one can search for $\tilde\Delta$ in the form
$\tilde\Delta=p^-\tilde\Delta_1 p^-$ and suppose that
\begin{equation}
p^+\tilde\Delta=0.
\end{equation}
Considering now Eq. (2.12) with $p^{ij}$ being presented in the form (2.9)
and using Eq. (2.13) one can get (after straightforward but extremely
tedious calculations) that
\begin{eqnarray}
\lefteqn{(-1)^{\epsilon_j\epsilon_k}p^{kl}{\stackrel\to\pa}_lp^{ij}+
\mbox{cycle}(ijk)=}\cr
&&-(-1)^{\epsilon_j\epsilon_k+\epsilon_l(\epsilon_i+\epsilon_\alpha)}
[\omega^{kl}-\omega^{ks}{\stackrel\to\pa}_s\chi^-_\rho
\tilde\Delta^{\rho\gamma}{\stackrel\leftarrow\pa}_p\chi^-_\gamma
\omega^{pl}]\omega^{in}{\stackrel\to\pa}_n\chi^-_\alpha[p^+_\delta{}^\alpha
{\stackrel\to\pa}_l\tilde\Delta^{\beta\delta}+{}\cr
&&{}+{\stackrel\to\pa}_l\tilde\Delta^{\alpha\delta}p^+_\delta{}^\beta]
{\stackrel\leftarrow\pa}_m\chi^-_\beta\omega^{mj}+\mbox{cycle}(ijk).
\end{eqnarray}
Since
\[ p^+_\delta{}^\alpha{\stackrel\to\pa}_n\chi^-_\alpha\equiv
-({\stackrel\to\pa}_np^+_\delta{}^\alpha)\chi^-_\alpha \]
the graded cycle (2.18) weakly vanishes and, consequently, the bracket (2.7)
possesses the Jacobi identity on the second class constraints surface
$\chi^-_\alpha\approx0$.

Thus, for building the generalized Dirac bracket, which is consistent
with setting all the (reducible) second class constraints strongly to
zero it is sufficiently to find a solution $p^\pm_\alpha{}^\beta$,
$\tilde\Delta^{\alpha\beta}$ of Eqs. (2.1), (2.3), (2.5), and (2.14)
which is compatible with Eqs. (2.6) and (2.17). Note that although the
proposed bracket allows correct elimination of the second class
constraints, some of its properties (see Eqs. (2.8), (2.10), (2.11),
and (2.18)) are fulfilled in a weak sense only. Reducibility of the
second class constraints, thus, manifests itself in weakening some
properties of the standard Dirac bracket.

\section{Applications}
\subsection{{\boldmath$N=1$} Brink--Schwarz superparticle}

As an example of a theory for which the proposed bracket construction
requires an extension of the initial phase space we consider $N=1$
Brink--Schwarz superparticle. The dynamics of the model in the original
phase space is determined by the following (first-order formalism) action [5]:
\begin{equation}
S=\int d\tau\, \left(p_m\Pi^m-\frac{ep^2}{2}\right),
\end{equation}
\[ \Pi^m\equiv\dot x^m-i\theta\Gamma^m\dot\theta. \]
For definiteness we consider ten-dimensional case here. The results of
the section, however, can be directly generalized to the case of other
dimensions. We use (generalized Majorana) notations, in which
$\theta^\alpha$ is a real M-W spinor $\alpha=1,\dots,16$, Dirac
matrices $\Gamma^m_{\alpha\beta}$ and $\tilde\Gamma^{m\,\alpha\beta}$
are real, symmetric, obeying the standard algebra
$\Gamma^m\tilde\Gamma^n+\Gamma^n\tilde\Gamma^m=2\eta^{mn}$.

Passing to the Hamiltonian formalism one can find the following
constraints for the model:
$$
p_m-\pi_m\approx0, \qquad \pi_{pm}\approx 0,
\eqno{(3.2a)}$$
$$
p_e\approx 0, \qquad \pi^2\approx 0,
\eqno{(3.2b)}$$
$$
\chi\equiv p_\theta-i\theta\Gamma^m\pi_m\approx 0
\eqno{(3.2c)}$$
where $(p_e,\pi_{pm},\pi_m,p_{\theta^\alpha})$ are momenta conjugate to
variables $(e,p^m,x^m,\theta^\alpha)$ respectively. The constraints
(3.2a) are second class and imply that the pair $(p,\pi_p)$ is pure
gauge. The constraints (3.2b) are first class. Among the spinor
constraints (3.2c), there are eight of first class and eight of second
class as a consequence of the equations (we use the Poisson bracket of the
form $\{x^n,\pi_m\}={\delta^n}_m$, $\{\theta^\alpha,p_{\theta\beta}\}=
-{\delta^\alpha}_\beta$)
$$
\{\chi_\alpha,\chi_\beta\}=2i(\Gamma^m\pi_m)_{\alpha\beta},
\eqno{(3.3a)}$$
$$
\Gamma^m\pi_m\tilde\Gamma^n\pi_n=\pi^2\approx0.
\eqno{(3.3b)}$$
\addtocounter{equation}{2}
Our aim now is to construct the projectors (2.1), (2.3) and the
generalized Dirac bracket (2.7) for the model. Technically, the task of
finding the needed projectors consists in building the matrix
${K_\alpha}^\beta$ satisfying the equations $K^2=1$,
$\Gamma^m\pi_mK\approx-\Gamma^m\pi_m$. One can show, however, that in the
initial phase space it is impossible to built the covariant object of such
a kind.

To construct the needed quantity let us introduce an additional vector
variable
$A_m$ and consider several trivial consequences of Eq. (3.3b)
\begin{eqnarray}
&&\Gamma^m\pi_m\tilde\Gamma^n\pi_n\Gamma^rA_r=
\pi^2\Gamma^rA_r\approx0,\cr
&&\Gamma^m\pi_m\tilde\Gamma^nA_n\Gamma^r\pi_r=2(\pi A)\Gamma^r\pi_r-
\Gamma^nA_n\pi^2\approx2(\pi A)\Gamma^r\pi_r,\\
&&\Gamma^m\pi_m\displaystyle\frac{1}{2}\Big(1+\frac{1}{2(\pi A)}
\tilde\Gamma^{[n}\Gamma^{r]}\pi_nA_r\Big)\equiv\Gamma^m\pi_m\tilde
p^+\approx 0.\nonumber
\end{eqnarray}
In agreement with Eq. (2.3),  the arising operator $\tilde p^+$ is a weak
eigenvector corresponding to zero eigenvalue for the matrix
$\Delta^*=2i\Gamma^m\pi_m$ and moreover it proves to be a weak
projector, i.e.,
\begin{equation}
\tilde p^{+2}\approx\tilde p^+.
\end{equation}
This equation can further be strengthened to yield a strict equality and
the results are
\begin{equation}
\begin{array}{c} p^+=\displaystyle\frac{1}{2}(1+K), \qquad
p^-=\frac{1}{2}(1-K),\\
K=\displaystyle\frac{1}{2\sqrt{(\pi A)^2-\pi^2A^2}}\tilde\Gamma^{[n}
\Gamma^{m]}\pi_nA_m, \qquad K^2=1.\end{array}
\end{equation}
Thus, to construct the needed projector operators it is sufficiently to
enlarge the initial phase space by means of adding one vector variable
$A_m$ only. To be consistent, we should then include this variable
into the original Lagrangian (3.1) (in a pure gauge manner). The
following action:
\begin{equation}
S=\int d\tau\,p_m(A^m-i\theta\Gamma^m\dot\theta)+B_m(A^m-\dot x^m)-
\frac{ep^2}{2}
\end{equation}
turns out to be suitable for this goal. In constructing this action we
were enforced to introduce one more additional variable $B_m$. Taking
into account the equations of motion for the new variables
\begin{equation}
\frac{\delta S}{\delta A^m}=p_m+B_m=0, \qquad
\frac{\delta S}{\delta B^m}=A_m-\dot x_m=0
\end{equation}
it is easy to show that the model (3.7) is on-shell equivalent to the
original superparticle (3.1).

The supersymmetry transformations are written now as
\begin{equation}
\delta\theta=\epsilon, \qquad \delta x^m=i\epsilon\Gamma^m\theta,
\qquad \delta A^m=i\epsilon\Gamma^m\dot\theta.
\end{equation}
Local $\alpha$- and $k$-symmetries [5, 24] take the form
\begin{eqnarray}
\delta_\alpha x^m=\alpha\dot x^m, &\qquad
\delta_\alpha\theta=\alpha\dot\theta^\beta,\cr
\delta_\alpha p^m=\alpha\dot p^m, &\qquad
\delta_\alpha e=(\alpha e)^\cdot,\\
\delta_\alpha B^m=\alpha\dot B^m, &\qquad
\delta_\alpha A^m=(\alpha A^m)^\cdot;\nonumber
\end{eqnarray}
\begin{equation}
\begin{array}{ll} \delta_k\theta=\tilde\Gamma^mp_mk, &\qquad
\delta_kx^m=i\theta\Gamma^m\delta\theta,\\
\delta_ke=4i\dot\theta k, &\qquad \delta_kA^m=(i\theta\Gamma^m
\delta\theta)^\cdot.\end{array}
\end{equation}
A complete constraint system of the model in the Hamiltonian formalism
is
$$
T_A:\qquad p_e\approx0, \qquad \pi^2\approx0;
\eqno{(3.12a)}$$
$$
\chi_\alpha:\qquad p_\theta-i\theta\Gamma^m\pi_m\approx 0;
\eqno{(3.12b)}$$
$$
\varphi_{A_1}:\qquad \pi_{Am}\approx 0;\qquad\qquad
\eqno{(3.12c)}$$
$$
\qquad\qquad\psi_{A_2}:\qquad \begin{array}[t]{ll} p_m-\pi_m\approx0, &
\pi_{pn}\approx 0,\\
p_m+B_m\approx 0, & \pi_{Bn}\approx 0\end{array}
\eqno{(3.12d)}$$
\stepcounter{equation}
where we denoted the momenta conjugate to variables $(e,x^m,p^m,
\theta^\alpha,A^m,B^m)$ as \linebreak $(p_e,\pi_m,\pi_{pm},p_{\theta\alpha},
\pi_{Am},\pi_{Bm})$ respectively. The constraints (3.12a), (3.12c)
are first class. Constraint system (3.12d) is second class. Among the
fermionic constraints (3.12b), half is first class and another half is
second class. Note that in the gauge $A^m\approx 0$ the constraint
system (3.12) precisely coincides with the Brink--Schwarz one [2, 5].

Using now the projectors (3.6) to split the spinor constraints into
redundant first and second class pieces $\chi^+\equiv\chi p^+$,
$\chi^-\equiv\chi p^-$
\begin{equation}
\begin{array}{l} \{\chi^+_\alpha,\chi^+_\beta\}\approx 0, \qquad
\{\chi^+_\alpha,\chi^-_\beta\}\approx 0,\\
\{\chi^-_\alpha,\chi^-_\beta\}=2i(p^-\Gamma^m\pi_mp^-)_{\alpha\beta}
\equiv\Delta_{\alpha\beta},\\
\{\chi^\pm_\alpha,\varphi_{A_1}\}\approx 0, \qquad
\{\chi^\pm_\alpha,\psi_{A_2}\}=0,\\
\{\chi^\pm_\alpha,T_A\}=0\end{array}
\end{equation}
and converting the matrix $\Delta_{\alpha\beta}$ in accordance with Eq.
(2.5)
\begin{equation}
\tilde\Delta^{\alpha\beta}=\frac{p^{-\alpha}{}_\delta(\tilde\Gamma^n
A_n)^{\delta\sigma}p^{-\beta}{}_\sigma}{2i\big(\sqrt{(\pi A)^2-
\pi^2A^2}+(\pi A)\big)}, \qquad \tilde\Delta\Delta=p^-
\end{equation}
one can write down the final expression for the generalized Dirac
bracket
\begin{eqnarray}
\lefteqn{\{A,C\}_D=\{A,C\}-
\{A,\chi^-_\alpha\}\frac{(p^-\tilde\Gamma^nA_np^-)^{\alpha\beta}}
{2i\big(\sqrt{(\pi A)^2-\pi^2A^2}+(\pi A)\big)}\{\chi^-_\beta,C\}-}\cr
&-\{A,{\pi_p}^m\}\{p_m-\pi_m,C\}+\{A,p^m-\pi^m\}\{\pi_{pm},C\}-
\{A,{\pi_B}^m\}\{p_m+B_m,C\}+{}\cr
&+\{A,p^m+B^m\}\{\pi_{Bm},C\}-\{A,p^m-\pi^m\}\{\pi_{Bm},C\}
+\{A,{\pi_B}^m\}\{p_m-\pi_m,C\}.
\end{eqnarray}
Since the projectors (3.6) satisfy the Eqs. (2.14), the
proposed bracket possesses (in a weak sense) all the rank and
algebraic properties of the standard Dirac bracket. Consistent covariant
elimination of the (reducible) second class constraints is now
possible. The explicit form of the fundamental phase variable brackets is
(we omit here the brackets corresponding to unphysical variables $(p,\pi_p)$,
$(B,\pi_B)$)
\begin{eqnarray}
&& \{\theta^\alpha,\theta^\beta\}_D=\displaystyle
\frac{i}{q}(p^-\nta p^-)^{\alpha\beta},\cr
&& \{\theta^\alpha,p_{\theta_\beta}\}_D=-{\delta^\alpha}_\beta+
\displaystyle\frac 1 2 p^{-\alpha}{}_\beta,\cr
&& \{p_{\theta_\alpha},p_{\theta_\beta}\}_D=-\displaystyle\frac i 2
(p^-\not\!\pi p^-)_{\alpha\beta},\cr
&& {} \cr
&& \{x^m,\theta^\alpha\}_D=-\displaystyle\frac{1}{q}
(\theta\Gamma^m\nta p^-)^\alpha-\frac iq
\Big(\chi^+\Big[\frac{\pa}{\pa\pi_m}p^-\Big]\nta p^-\Big)^\alpha,\cr
&&
\{x^m,p_{\theta_\alpha}\}_D=\displaystyle\frac{i}{2}(\theta\Gamma^mp^-)_\alpha
-\frac 12\Big(\chi^+\frac{\pa}{\pa\pi_m}p^-\Big)_\alpha,\cr
&& \{{\pi_A}^m,\theta^\alpha\}_D=\displaystyle\frac iq
\Big(\chi^+\Big[\frac{\pa}{\pa A_m}p^-\Big]\nta p^-\Big)^\alpha,\cr
&& \{{\pi_A}^m,p_{\theta_\alpha}\}_D=\displaystyle\frac 12
\Big(\chi^+\frac{\pa}{\pa A_m}p^-\Big)_\alpha,\\[2ex]
&& \{x^m,x^n\}_D=\displaystyle\frac iq \theta\Gamma^mp^-\nta p^-\Gamma^n\theta-
\frac 1q \theta\Gamma^m\nta\Big[\frac{\pa p^-}{\pa\pi_n}\Big]\chi^++{}\cr
&& \qquad {}+\frac 1q \theta\Gamma^n\nta\Big[\frac{\pa
p^-}{\pa\pi_m}\Big]\chi^+
-\frac iq \chi^+\Big[\frac{\pa p^-}{\pa\pi_m}\Big]\nta\Big[
\frac{\pa p^-}{\pa\pi_n}\Big]\chi^+,\cr
&& \{x^m,\pi_n\}_D={\delta^m}_n,\cr
&& \{x^m,\pi_{A_n}\}_D=\displaystyle\frac 1q \theta\Gamma^m\nta\Big[
\frac{\pa p^-}{\pa A^n}\Big]\chi^++\frac iq \chi^+\Big[\frac{\pa p^-}
{\pa\pi_m}\Big]\nta\Big[\frac{\pa p^-}{\pa A^n}\Big]\chi^+,\cr
&& \{A^n,\pi_{A_m}\}_D={\delta^n}_m,\cr
&& \{\pi_{A_m},\pi_{A_n}\}_D=-\displaystyle\frac iq \chi^+\Big[\frac{\pa p^-}
{\pa A^m}\Big]\nta\Big[\frac{\pa p^-}{\pa A^n}\Big]\chi^+,\nonumber
\end{eqnarray}
were we denoted $q=2\big(\sqrt{(\pi A)^2-\pi^2A^2}+(\pi A)\big)$,
${\nta}=\tilde\Gamma^nA_n$, $\not\!\!\pi=\Gamma^n\pi_n$ and used the
identities $p^-\not\!\pi=\not\!\pi p^-$, $p^-\nta=\nta p^-$,
$p^+\not\!\pi p^-=p^+\nta p^-=0$, $p^-\pa p^-p^-=0$.

Note as well that the presented scheme allows the covariant
(redundant) gauge
\begin{equation}
\theta^+\equiv p^+\theta\approx 0
\end{equation}
for the fermionic first class constraints.

Let us briefly discuss a relation between the considered formulation
and the Hamiltonian null-vectors approach of Ref. 2.

The basic idea of the construction proposed in Ref. 2 was to introduce
two null vectors
\begin{equation}
n^2=0, \qquad r^2=0,\qquad nr=-1
\end{equation}
(which were considered as pure gauge variables) to separate
the initial fermionic second class constraints in covariant and
redundant way
\begin{equation}
\psi\equiv\chi\not{\tilde n}\not r\approx 0.
\end{equation}
Note that in the presence of the constraints (3.18) the operators
\begin{equation}
\tilde p^-=\frac{1}{2(nr)}\not{\tilde n}\not r, \qquad
\tilde p^+=\frac{1}{2(nr)}\not{\tilde r}\not n
\end{equation}
form weak projectors (after constructing the Dirac bracket associated
with the full system of second class constraints [2], the constraints
(3.18) can be considered as strong equations and the operators (3.20) become
strong projectors). One can believe, therefore, that the new variables
were introduced to construct a projector operator extracting the fermionic
second class constraints.

Return now to the formulation (3.7) and let us use the variables $A^m$,
$\pi^n$ to define a pair of strong null vectors (see also Ref. 25)
\begin{equation}
\begin{array}{l} {n'}^m=\displaystyle\frac{1}{c}\big[A^2\pi^m-\big(
(A\pi)+\sqrt{(A\pi)^2-A^2\pi^2}\big)A^m\big],\\
{r'}^m=\displaystyle\frac{1}{c}\big[A^2\pi^m-\big(
(A\pi)-\sqrt{(A\pi)^2-A^2\pi^2}\big)A^m\big],\end{array}
\end{equation}
\[ n^{'2}\equiv 0, \qquad r^{'2}\equiv 0, \qquad n'r'\equiv -1 \]
where we denoted $c=\sqrt{2A^2((\pi A)^2-\pi^2A^2)}$. The crucial
observation is that the following identities:
\begin{equation}
\begin{array}{l} p^-=\displaystyle\frac{1}{2}\Big(1-\frac{1}{2b}
\tilde\Gamma^{[n}\Gamma^{m]}\pi_nA_m\Big)\equiv\frac{1}{2(n'r')}
\not{\tilde n}'\not r',\\
p^+=\displaystyle\frac{1}{2}\Big(1+\frac{1}{2b}\tilde\Gamma^{[n}
\Gamma^{m]}\pi_nA_m\Big)\equiv\frac{1}{2(n'r')}\not{\tilde r}'
\not n',\end{array}
\end{equation}
where $b\equiv\sqrt{(\pi A)^2-\pi^2A^2}$, are fulfilled.

Thus, the basic constructions of Ref. 2 can be reproduced within the
context of the theory (3.7) and, in this sense, there is a correspondence
between two formulations. It should be noted, however, that the model
(3.7) is free from some difficulties of Ref. 2. In particular, the
constraint system for additional variables ($\varphi_{A_1}$, $\psi_{A_2}$
in terminology of Sec. 2) in Ref. 2 is reducible and very complicated
as compared to Eqs.  (3.12c) and (3.12d). The operator separating the
(redundant) first class constraints $(\Gamma^m\pi_m)$ is not a projector
and a covariant proof of the equivalence between the splitted and original
mixed constraints presents a special problem. In our case this proof is
evident. Note as well that the formulation of Ref. 2 is essentially
Hamiltonian.

\subsection{{\boldmath$N=1,D=9$} massive superparticle with Wess--Zumino term}

In this subsection, as an example of the model for which the
generalized bracket can be constructed in the initial phase space we
consider $N=1,D=9$ massive superparticles with Wess--Zumino term [19].

The basic observation lies in the fact that in certain dimensions
there exists Lorentz invariant, real, symmetric tensor
$X_{\alpha\beta}$ (in addition to the Dirac matrices) which can be used
to build the needed projectors. To prove the existence of such a tensor
for the case concerned, let us construct the minimal spinor
representation of $SO(1,8)$ (which has a complex dimension $2^{(D-1)/2}=16$)
and the corresponding $\Gamma$-matrices in the explicit form.

For this aim it is sufficiently to find nine $16\times16$ matrices
${\Gamma^m}_\alpha{}^\beta$ satisfying the equation
$\Gamma^m\Gamma^n+\Gamma^n\Gamma^m=-2\eta^{mn}$, $m=0,1,\dots,8$.
Taking into account that $SO(1,9)$-matrices from the previous section
(which we denote now as ${\gamma^m}_{\alpha\beta}$,
$\tilde\gamma^{m\alpha\beta}$) have the needed dimension, one can consider
the following decomposition:
\begin{equation}
\begin{array}{c} X_{\alpha\beta}\equiv{\gamma^9}_{\alpha\beta}, \qquad
\tilde X^{\alpha\beta}\equiv\tilde\gamma^{9\alpha\beta}, \qquad
{\Gamma^m}_{\alpha\beta}\equiv{\gamma^m}_{\alpha\beta},\\
\tilde\Gamma^{m\alpha\beta}\equiv\tilde\gamma^{m\alpha\beta}, \qquad
{\Gamma^m}_\alpha{}^\beta\equiv{\Gamma^m}_{\alpha\delta}\tilde
X^{\delta\beta},
\qquad m=0,1,\dots,8.\end{array}
\end{equation}
The properties of $\gamma^m$, $\tilde\gamma^m$ induce the following relations
for $X$ and $\Gamma$:
\begin{eqnarray}
&& X_{\alpha\beta}=X_{\beta\alpha}, \qquad X^*_{\alpha\beta}=X_{\alpha\beta},
\qquad X_{\alpha\beta}\tilde X^{\beta\gamma}={\delta_\alpha}^\gamma,\cr
&& X_{\alpha\beta}\tilde\Gamma^{m\beta\gamma}+{\Gamma^m}_{\alpha\beta}
\tilde X^{\beta\gamma}=0,\\
&& {\Gamma^m}_\alpha{}^\beta{\Gamma^n}_\beta{}^\gamma+
{\Gamma^n}_\alpha{}^\beta{\Gamma^m}_\beta{}^\gamma=-2\eta^{mn}.\nonumber
\end{eqnarray}
Thus, the minimal spinor representation of $SO(1,8)$ is a complex spinor
$\psi^\alpha$ transforming, by definition, as follows
\begin{equation}
\delta\psi^\alpha=\frac 12 \omega_{mn}
(\tilde\Gamma^{mn})^\beta{}_\alpha\psi^\alpha,
\end{equation}
where
\[ \tilde\Gamma^{mn}=\frac 14 (\tilde\Gamma^m\Gamma^n-
\tilde\Gamma^n\Gamma^m). \]
Since the combination $\bar\psi_\alpha\equiv X_{\alpha\beta}\psi^\beta$
is transformed as $\delta\bar\psi_\alpha=-\frac 12 \omega_{mn}
(\bar\psi\tilde\Gamma^{mn})_\alpha$, we conclude that the expression
$(\psi^\alpha X_{\alpha\beta}\varphi^\beta)$ is a scalar, i.e.,
$X_{\alpha\beta}$ is $SO(1,8)$-invariant matrix. Note that the reality
condition ${\psi_\alpha}^*=\psi_\alpha$ is consistent with this construction.

The action functional of the theory is given by the expression
\begin{equation}
S=\int d\tau\, \Big(e^{-1}\frac{\Pi^2}{2}-\frac{m^2e}{2}+im\dot\theta
X\theta\Big),
\end{equation}
\[ \Pi^m=\dot x^m-i\theta\Gamma^m\dot\theta, \]
where we have introduced the einbein tangent to the superparticle
wordline as opposed to the action of Ref. 19. In this formulation a
form of the local symmetries becomes evident
$$
\delta_\alpha x^m=\alpha\dot x^m, \qquad \delta_\alpha\theta^\beta=
\alpha\dot\theta^\beta, \qquad \delta_\alpha e=(\alpha e)^\cdot;
\eqno{(3.27a)}$$
$$
\delta_k\theta=(\tilde\Gamma^m\Pi_m+me\tilde X)k, \qquad
\delta_kx^m=i\theta\Gamma^m\delta\theta, \qquad \delta_k e =
4ie\dot\theta k,
\eqno{(3.27b)}$$
\stepcounter{equation}
The constraint system of the theory in the Hamiltonian formalism is
\begin{equation}
p_e\approx0, \qquad \pi^2+m^2\approx0, \qquad \chi\equiv p_\theta-i\theta
(\Gamma^n\pi_n+mX)\approx0,
\end{equation}
where the variables $(p_e,\pi_m, p_{\theta_\alpha})$ are momenta conjugate
to $(e,x^m,\theta^\alpha)$ respectively. Dynamics of the model is governed
by the Hamiltonian
\begin{equation}
H=p_e\lambda_e+\lambda_\theta\chi+\frac{e}{2}(\pi^2+m^2)
\end{equation}
where $\lambda_e,\lambda_\theta$ are Lagrange multipliers to the
constraints $p_e$ and $\chi$ respectively. The bosonic constraints in
Eq. (3.28) are first class, while there are half of first class
constraints and half of second class ones among $\chi_\alpha$
$$
\{\chi_\alpha,\chi_\beta\}=2i(\Gamma^n\pi_n+mX)_{\alpha\beta},
\eqno{(3.30a)}$$
$$
(\Gamma^m\pi_m+mX)(\tilde\Gamma^n\pi_n+m\tilde X)=m^2+\pi^2\approx0.
\eqno{(3.30b)}$$
\stepcounter{equation}
Let us construct the generalized Dirac bracket for the model. The first
step is building the projector operators satisfying Eqs. (2.1), (2.3),
and (2.14). Taking into account Eqs. (3.30b) and (3.24) one can find
the needed quantities (see also Ref. 19)
\begin{equation}
p^+=\frac{1}{2}(1+K), \quad p^-=\frac{1}{2}(1-K), \quad
{K_\alpha}^\beta=\frac{1}{\sqrt{-\pi^2}}\pi_m(X\tilde\Gamma^m)
_\alpha{}^\beta, \quad K^2=1.
\end{equation}
It is straightforward to check as well that the following identities:
$$
p^+\Big(X+\frac{1}{\sqrt{-\pi^2}}\Gamma^m\pi_m\Big)=0, \qquad
p^-\Big(X-\frac{1}{\sqrt{-\pi^2}}\Gamma^m\pi_m\Big)=0;
\eqno{(3.32a)}$$
$$
p^\pm\Gamma^m\pi_m=\Gamma^m\pi_mp^\pm, \qquad p^\pm X=Xp^\pm,
\eqno{(3.32b)}$$
$$
p^+\Gamma^m\pi_mp^-=0, \qquad p^+Xp^-=0
\eqno{(3.32c)}$$
are fulfilled.
\stepcounter{equation}

In the presence of the projectors the fermionic constraints are
splitted into (redundant) first and second class pieces $\chi^+ \equiv
p^+\chi$ and $\chi^-\equiv p^-\chi$:
\begin{eqnarray}
&& \{\chi^+_\alpha,\chi^+_\beta\}=\displaystyle
\frac{2i}{\sqrt{-\pi^2}}(\sqrt{-\pi^2}-m)(p^+\Gamma^n\pi_n)_{\alpha\beta}
\approx0,\cr
&& \{\chi^+_\alpha,\chi^-_\beta\}=0,\\
&& \{\chi^-_\alpha,\chi^-_\beta\}=\displaystyle\frac{2i}{\sqrt{-\pi^2}}
(\sqrt{-\pi^2}+m)(p^-\Gamma^n\pi_n)_{\alpha\beta}\nonumber
\end{eqnarray}
($\pi^m$ is supposed to be a space-like vector, therefore the constraint
$\pi^2+m^2\approx0$ is equivalent to $m-\sqrt{-\pi^2}\approx0$). Using
Eqs. (3.32) one can choose, further, more simple basis of the
constraints:
\begin{equation}
p^+_\theta\approx0,\quad \chi^-\equiv p^-_\theta-i\theta^-X(m+
\sqrt{-\pi^2})\approx0, \quad m-\sqrt{-\pi^2}\approx 0, \quad
p_e\approx 0
\end{equation}
where we denoted $p^\pm_\theta=p^\pm p_\theta$,
$\theta^\pm=p^\pm\theta$. In this representation, finding
$\tilde\Delta^{\alpha\beta}$ for Eq. (2.5) and building the generalized
bracket present no a special problem. The results are
$$
\{A,B\}_D=\{A,B\}-\{A,\chi^-_\alpha\}\tilde\Delta^{\alpha\beta}
\{\chi^-_\beta,B\},
\eqno{(3.35a)}$$
$$
\tilde\Delta^{\alpha\beta}=\frac{(\tilde Xp^-)^{\alpha\beta}}
{2i(m+\sqrt{-\pi^2})}.
\eqno{(3.35b)}$$
\stepcounter{equation}
Since the projectors (3.31) satisfy Eq. (2.14) the constructed bracket
possesses (in a weak sense) all the rank and algebraic properties of the
standard Dirac bracket. Thus, Eqs. (3.31), (3.35a), and (3.35b) specify
the generalized Dirac bracket for the massive superparticle with
Wess--Zumino term.

The explicit form of the fundamental phase variable brackets is
\begin{eqnarray}
&& \{\theta^\alpha,\theta^\beta\}_D=\displaystyle
\frac{i}{2(m+\sqrt{-\pi^2})}(\tilde Xp^-)^{\alpha\beta},\cr
&& \{\theta^\alpha,p_{\theta_\beta}\}_D=-{\delta_\beta}^\alpha+
\displaystyle\frac 1 2 p_\beta^{-\alpha},\cr
&& \{p_{\theta_\alpha},p_{\theta_\beta}\}_D=-\displaystyle\frac i 2
(m+\sqrt{-\pi^2})(Xp^-)_{\alpha\beta},\cr\vspace{2ex}
&&
\{x^m,\theta^\alpha\}_D=\displaystyle\frac{i}{4(m+\sqrt{-\pi^2})\sqrt{-\pi^2}}
\Big(\tilde\Gamma^m+\frac{\pi^m\tilde X}{\sqrt{-\pi^2}}\Big)^{\alpha\lambda}
\big(p_\theta^+-i\theta^+X(m+\sqrt{-\pi^2})\big)_\lambda+{}\cr
&& \qquad{}+\frac{1}{2(m+\sqrt{-\pi^2})\sqrt{-\pi^2}}\pi^m\theta^{-\alpha},\\
&& \{x^m,p_{\theta_\alpha}\}_D=\displaystyle\frac{1}{4\sqrt{-\pi^2}}
\Big(X\tilde\Gamma^m+\frac{\pi^m}{\sqrt{-\pi^2}}\Big)
\big(p_\theta^+-i\theta^+X(m+\sqrt{-\pi^2})\big)_\alpha-
\frac{i}{2\sqrt{-\pi^2}}\pi^m(\theta^-X)_\alpha,\cr\vspace{2ex}
&& \{x^m,\pi_n\}_D={\delta^m}_n,\cr
&& \{x^m,x^n\}_D=\displaystyle\frac{i}{16(m+\sqrt{-\pi^2})\pi^2}\big(
p_\theta^+-i\theta^+X(m+\sqrt{-\pi^2})\big)\tilde\Gamma^{[m}X
\tilde\Gamma^{n]}\big(p_\theta^+-i\theta^+X(m+\sqrt{-\pi^2})\big)-\cr
&& \qquad{}-\displaystyle\frac{1}{4(m+\sqrt{-\pi^2})\pi^2}\theta^-X
\tilde\Gamma^{[m}\pi^{n]}\big(p_\theta^+-i\theta^+X(m+\sqrt{-\pi^2})\big),\nonumber
\end{eqnarray}
were we denoted $A^{[n}B^{m]}\equiv A^nB^m-A^mB^n$ and used the
identity $\Big(\tilde X-\frac{1}{\sqrt{-\pi^2}}\tilde\Gamma^n\pi_n\Big)
p^+\equiv0$.

The following remarks seem to be relevant: First, taking into account
Eq. (2.18) one can check that the strong Jacobi identity problem
appears only in the cycles including the variable $x^m$. Secondly, the
considered scheme admits the covariant (reducible) gauge
\begin{equation}
\theta^+\approx0
\end{equation}
and the corresponding Dirac-type bracket
\begin{equation}
\{A,B\}_D=\{A,B\}+\{A,p^+_\theta\}p^+\{\theta^+,B\}+\{A,\theta^+\}p^+
\{p_\theta^+,B\}-\{A,\chi^-\}\tilde\Delta\{\chi^-,B\}.
\end{equation}
Thus, the physical sector of the model is exhausted by the variables
$(x^m,\pi_n,\theta^{-\alpha})$ with commutation relations being
presented in the form
\begin{equation}
\begin{array}{l} \{\theta^{-\alpha},\theta^{-\beta}\}=-\displaystyle
\frac{1}{2i(m+\sqrt{-\pi^2})}(p^-\tilde X)^{\alpha\beta},\\
\{\theta^{-\alpha},x^m\}=\displaystyle\frac 1{2\sqrt{-\pi^2}}
(\theta^-X\tilde\Gamma^m)^\alpha+\frac 12\biggl(\frac 1{\pi^2}-
\frac 1{(m+\sqrt{-\pi^2})\sqrt{-\pi^2}}\biggr)\pi^m\theta^{-\alpha},\\
\{x^m,\pi_n\}={\delta^m}_n,\\
\{x^m,x^n\}=-\displaystyle\frac{i(m+\sqrt{-\pi^2})}{4\pi^2}
\theta^-\Gamma^{[m}\tilde\Gamma^{n]}X\theta^-.\end{array}
\end{equation}
It is straightforward to check now that the Jacobi identity for the
brackets (3.39) is fulfilled in a strong sense. Thirdly, it was shown in
Refs. 19 and 26 that appearing the Wess--Zumino term in the superparticle
action plays a role of introducing a central charge into the super-Poincar\'e
algebra. As was seen above, it was this quantity which allowed to
construct the generalized Dirac bracket for the model.

\section{Off-shell continuation of the generalized \protect\newline brackets
and the unified constraint dynamics}

In constructing the generalized Dirac bracket associated with the
infinitely reducible second class constraints, the essential property
which was embedded into the scheme was the weak Jacobi identity
(the special restrictions were to be imposed to provide the property).
To apply standard covariant quantization methods in a {\it full} phase
space, it is necessary then to continue the bracket up to one with
the strong Jacobi identity. Note in this context that the bracket
structure is a sum of its body on the constraint surface, its soul
(fermionic terms) and the constraints involved. The rank of the bracket
is defined by the first terms only. One can believe that the strong
Jacobi identity is absent because not all needed fermions and constraints
were added to the body. We may add arbitrary (the most general) combination
of such terms with some coefficient functions. Then the requirement of
the strong Jacobi identity for the new bracket will fix these functions.

Another serious motivation for studying the question concerns a possibility
to apply the quantization scheme by Batalin and Tyutin [20--22] to the
superparticle models. The remarkable feature of the formalism developed
in Refs. 20--22 lies in the fact that it, in principle, allows to avoid
the explicit separation of the constraints into the first and second class
ones. Let us enumerate some relevant facts. A solution of quantum generating
equations (in the lowest orders) implies the following defining relations
for the classical counterparts [20]
\begin{eqnarray}
&& \{\Gamma^A,\Gamma^B\}=D^{AB}+Z^{AB}{}_\alpha\Theta^\alpha,\\
&& \{\Gamma^A,\Theta^\alpha\}=E^{A\alpha}+Y^{A\alpha}{}_\beta
\Theta^\beta,\\
&& \{\Theta^\alpha,\Theta^\beta\}=U^{\alpha\beta}{}_\gamma\Theta^\gamma,\\
&& \big(\{\Gamma^A,D^{BC}\}+Z^{AB}{}_\alpha E^{C\alpha}
(-1)^{\epsilon_\alpha\epsilon_C}\big)(-1)^{\epsilon_A\epsilon_C}+
\mbox{cycle}(ABC)=X^{ABC}{}_\alpha\Theta^\alpha,
\end{eqnarray}
where $\Gamma^A$ are variables and $\Theta^\alpha$ are linearly-independent
constraints of a theory. The numers $M'$ ($M''$) of first (second) class
constraints among $\Theta^\alpha$, $\alpha=1,\dots,M'+M''$, are fixed by
conditions
\begin{equation}
\rank \|E^{A\alpha}\|\big|_{\Theta=0}=M',\quad
{\rm corank}\,\|D^{AB}\|\big|_{\Theta=0}=M''.
\end{equation}
The quantities $D^{AB}$, $E^{A\alpha}$ must be embedded into the scheme
from the outset as a boundary conditions for generating equations and, as
it seen from Eqs. (4.3), (4.4), satisfy to weakened version of the standard
properties of the Dirac bracket.

In Refs. 20--22 an existence of a solution of quantum generating equations
with such boundary conditions was shown. In particular, there exists, in
principle, the quantum analogue of $\hat Z^{AB}{}_\alpha$ supplying the
Jacobi identity for fundamental brackets (4.1). In previous sections we found
the quantities $D$ and $E$ for the concrete models. To complete the scheme
at the classical level, we investigate the question of existence the
{\it covariant} quantity ${Z_\alpha}^{AB}$. Namely, for massive
superparticle considered above, it is possible, instead of continuation
of the available $D$ and $E$, to write down the most general Poincar\'e
covariant ansatz for fundamental brackets (4.1) with an accuracy of some
scalar coefficients. The coefficients of the body ansatz will be found
from the requirement that the conditions (4.1)--(4.3), (4.5) are fulfilled.
The remaining coefficients will then be fixed by demanding the strong Jacobi
identity for the bracket. Thus, our considerations are not related to an
existence of projectors or a particular form of the quantities $D$, $E$.

The basic observation lies in the fact that there arises a contradictory
system of equations for the coefficients, and our result looks as
follows: for the $D=9$ massive superparticle it is impossible to
construct a Poincar\'e covariant bracket obeying the conditions
(4.1)--(4.5).

To prove the fact let us demonstrate first the following assertion
(the fermionic variables and constraints from the subsec. 3.2 are denoted
now as $(\tilde\theta^\alpha,p_{\theta\alpha})\equiv Z^A$ and
$\tilde L_\alpha$, respectively):

{}From the conditions

a) the constraints $\tilde L_\alpha\approx0$, $T\equiv\pi^2+
m^2\approx0$ are in involution;

b) the rank conditions
\begin{equation}
{\rm corank}\,\{\tilde L_\alpha,\tilde L_\beta\}\big|_{L=T=0}=8, \qquad
\rank\{Z^A,\tilde L_\alpha\}\big|_{L=T=0}=8
\end{equation}
are fulfilled, it follows that a body of the bracket on the constraints
surface is determined in the odd-sector as
$$
\{L_\alpha,L_\beta\}=O^1(T)+O^1(L)O^1(\theta,L),
\eqno{(4.7a)}$$
$$
\{L_\alpha,\theta^\beta\}=-{p^+}_\alpha{}^\beta+O^1(T)
{p^-}_\alpha{}^\beta+O^1(T){p^+}_\alpha{}^\beta +O^2(\theta,L),
\eqno{(4.7b)}$$
$$
\{\theta^\alpha,\theta^\beta\}=-\frac{1}{4im}(\tilde Xp^-)^{\alpha\beta}+
O^1(T)(\tilde Xp^+)^{\alpha\beta}+O^1(T)(\tilde Xp^-)^{\alpha\beta}+
O^2(\theta,L).
\eqno{(4.7c)}$$
\stepcounter{equation}
where the brackets were written in terms of the shifted variables
$(\theta^\alpha,L_\alpha)$
\begin{eqnarray}
&& \theta^\alpha=\frac 1{2\sqrt{-imc_2}}\Big[\tilde\theta^\alpha -
\frac{c_1}{2k_2}\tilde X^{\alpha\beta}\big(p_{\theta\beta}-
i\tilde\theta^\gamma(\Gamma^m\pi_m+mX)_{\gamma\beta}\big)\Big],\cr\vspace{2ex}
&& L_\alpha=-\frac{2\sqrt{-imc_2}}{k_2}\big[p_{\theta\alpha}-
i\tilde\theta^\gamma(\Gamma^m\pi_m+mX)_{\gamma\alpha}\big],\\
&& c_2(\pi^m)\ne0, \qquad k_2(\pi^m)\ne0;\nonumber
\end{eqnarray}
and we used the weak projectors
\begin{equation}
{p^\pm}_\alpha{}^\beta\equiv\frac 12\Big(1\pm\frac 1m X\tilde\Gamma^m
\pi_m\Big)_\alpha{}^\beta.
\end{equation}
The terms proportional to $\theta$ and $L$ are denoted as
$O^1(\theta,L)$; analogously we denoted $O^2(\theta,L)\equiv\theta^2\dots+
\theta L\dots+L^2\dots$\,. The symbol $O^1(T)$ denotes the terms linear in
the bosonic constraint. The coefficients $k_i,c_i,a_i,\dots$, arising in all
expressions are scalar functions depending on the variable $\pi^m$ only.
Evidently, (non)existence of the bracket with the needed properties in terms
of shifted variables implies the same for the initial bracket.

To prove the assertion, note that under the condition a) the most
general Poincar\'e covariant ansatz for the brackets in the considered sector
can be put into the following form (where only the body on the constraint
surface is written in an explicit form)
$$
\{\tilde L_\alpha,\tilde L_\beta\}=O^1(T)+O^1(\tilde L)
O^1(\tilde\theta,\tilde L),
\eqno{(4.10a)}$$
$$
\{\tilde L_\alpha,\tilde\theta^\beta\}=k_1{p^-}_\alpha{}^\beta+
k_2{p^+}_\alpha{}^\beta+O^2(\tilde\theta,\tilde L)+
O^1(T){p^-}_\alpha{}^\beta+O^1(T){p^+}_\alpha{}^\beta,
\eqno{(4.10b)}$$
$$
\{\tilde\theta^\alpha,\tilde\theta^\beta\}= c_1\tilde Xp^++ c_2(\tilde
Xp^-)+O^2(\tilde\theta,\tilde L)+O^1(T).
\eqno{(4.10c)}$$
\stepcounter{equation}

To evaluate consequences of the rank conditions (4.6) let us pass to the
rest frame
\begin{eqnarray}
& \pi^m=(m,0,\dots,0), \qquad \pi^2=-m^2,\cr
& \Gamma^m\pi_m+mX=2m\left(\begin{array}{cc} 1 & 0\\0 & 0\end{array}
\right), \qquad \tilde\Gamma^m\pi_m+m\tilde X=-2m\left(
\begin{array}{cc} 0 & 0\\0 & 1\end{array}\right),\\
& p^-=\left(\begin{array}{cc} 1 & 0\\0 & 0\end{array}\right), \qquad
p^+=\left(\begin{array}{cc} 0 & 0\\0 & 1\end{array}\right).\nonumber
\end{eqnarray}
As a consequence of these relations, the body of the bracket
in the odd-sector (when restricted to the constraint surface) is
\begin{equation}
\{z^A,z^B\}=\begin{array}{cccccccccc}
& & & \tilde L & & & & \tilde\theta \\
& \vline &     & &  & | & k_1{\bf 1}_8 & \vdots & 0 & \vline\\
\tilde L & \vline & & 0 & & | & \cdots\cdots & \vdots & \cdots\cdots & \vline\\
& \vline & & & & | & 0 & \vdots & k_2{\bf 1}_8 & \vline\\
& \vline & - - - & - & - - - & | & - - - & - & - - - & \vline\\
& \vline & k_1{\bf 1}_8 & \vdots & 0 & | & & & & \vline\\
\tilde\theta & \vline & \cdots\cdots & \vdots & \cdots\cdots & | & & * & &
\vline\\
& \vline & 0 & \vdots & k_2{\bf 1}_8 & | & & & & \vline\end{array}.
\end{equation}
If $k_1\ne0$ and $k_2\ne0$ we have a nondegenerate matrix. So, to satisfy
Eq. (4.6) it is necessary to assume that $k_1=0$ (another possibility
$k_2=0$ can be considered along the same lines). Thus, instead
of (4.10b) we have
\begin{equation}
\{\tilde L_\alpha,\tilde\theta^\beta\}=k_2{p^+}_\alpha{}^\beta+
O^2(\tilde\theta,\tilde L)+O^1(T){p^-}_\alpha{}^\beta+
O^1(T){p^+}_\alpha{}^\beta.
\end{equation}
Shifting then $\tilde\theta^\alpha$
\begin{equation}
\tilde\theta^\alpha\to\tilde\theta^\alpha-\frac{c_1}{2k_2}
\tilde X^{\alpha\beta}\tilde L_\beta,
\end{equation}
one can get
\begin{equation}
\{\theta^\alpha,\theta^\beta\}=c_2(\tilde Xp^-)^{\alpha\beta}+
O^2(\theta,L)+O^1(T).
\end{equation}
Subsequent renormalizations of $\theta$ and $L$ can further be used
to eliminate the coefficients $k_2\ne0$ from the body of the bracket,
what will reproduce Eqs. (4.7a)--(4.7c), (4.8). The structure of brackets
does not change under all needed renormalizations since all the arising
additional contributions are of orders $O^2(\theta,L)$, $O^1(T)$. Note
that the body of the bracket (4.7) and the corresponding expressions from
(3.36) are the same.

Let us add the remaining variables and write the most general
Poincar\'e covariant ansatz for all brackets
$$
\begin{array}{l} \{\theta^\alpha,\theta^\beta\}=-\displaystyle
\underline{\frac{(\tilde Xp^-)^{\alpha\beta}}{4im}}+O^1(T)\tilde Xp^+O^1(T)
\tilde Xp^-+O^2(\theta,L)+\dots\\[1ex]
\{L_\beta,\theta^\alpha\}=\underline{-{p^+}_\beta{}^\alpha}+
O^1(T)p^-+O^1(T)p^+ +O^2(\theta,L)+\dots\\[1ex]
\{L_\alpha,L_\beta\}=O^1(T)+O^1(L)O^1(\theta,L)+\dots\end{array}
\qquad\qquad\qquad\vspace{6pt}
\eqno{(4.16a)}$$
$$
\begin{array}{l} \{x^m,\theta^\alpha\}=\underline{\theta^\delta\big[\Gamma^m
\tilde X(a_1p^++a_2p^-)}+\pi^m(\tilde{\tilde a}_1p^++\tilde{\tilde
a}_2p^-)\big]_\delta{}^\alpha+{}\\[1ex]
\qquad{}+\underline{L_\delta\big[\tilde\Gamma^m(b_1p^++b_2p^-)}+\pi^m\tilde X
(\tilde{\tilde b}_1p^++\tilde{\tilde b}_2p^-)
\big]^{\delta\alpha}+O^1(T)O^1(\theta,L)+O^3(\theta,L)+\dots\\[1ex]
\{x^m,L_\alpha\}=\underline{\theta^\delta\big[\Gamma^m (c_1p^++c_2p^-)}+\pi^mX
(\tilde{\tilde c}_1p^++\tilde{\tilde c}_2p^-)\big]_{\delta\alpha}+{}\\
\qquad{}+\underline{L_\delta\big[\Gamma^m\tilde X(d_1p^++d_2p^-)}+\pi^m
(\tilde{\tilde d}_1p^++\tilde{\tilde d}_2p^-)
\big]_\alpha{}^\delta+O^1(T)O^1(\theta,L)+O^3(\theta,L)+\dots\\
\{\pi^m,\theta^\alpha\}=O^1(T)+O^1(\theta,L)+\dots\\
\{\pi^m,L_\alpha\}=O^1(L)+O^1(T)O^1(\theta,L)+O^1(L)O^2(\theta,L)
+\dots\end{array}\quad\vspace{6pt}
\eqno{(4.16b)}$$
$$
\begin{array}{l} \{\pi^m,\pi^n\}=O^1(T)+O^2(\theta,L)+\dots\\
\{x^m,\pi_n\}=\underline{{\delta^m}_n}+O^1(T)+O^2(\theta,L)+\pi^m\pi_nG+\dots\\[1ex]
\{x^m,x^n\}=\underline{\displaystyle\frac 1 2\theta^\alpha\theta^\beta
[-\Gamma^{[m}\tilde X\Gamma^{n]}g_1+\dots]_{\beta\alpha}+
\theta^\alpha L_\beta[-\tilde\Gamma^{[m}\Gamma^{n]}g_3+\dots]^\beta
{}_\alpha}\\[1ex]
\qquad{}+\underline{\displaystyle\frac 1 2 L_\alpha L_\beta
[-\tilde\Gamma^{[m}\Gamma^{n]}\tilde X
h_1+\dots]^{\beta\alpha}}+O^1(T)+\dots\end{array}\qquad\qquad\vspace{6pt}
\eqno{(4.16c)}$$
\stepcounter{equation}
It is straightforward to check (by making use of Eqs. (3.32)) that the
coefficient matrices in Eqs. (4.16) are of the most general form.

Now, we may require the Jacobi identity for different cycles. For our
purposes, it proves to be sufficient to consider only four cycles
including the variables $(x^m,\theta^\alpha,L_\beta)$,
$(x^m,\theta^\alpha,\theta^\beta)$, $(x^m,L_\alpha,L_\beta)$, and
$(x^m,x^n,L_\alpha)$, and analyze the terms of zeroth and first orders
in $\theta$ and $L$ carrying the free vector indices $m,n$ on the
$\Gamma$-matrices only. All another contributions are neglected.
In particular, one can directly verify that only the stressed terms in
Eqs. (4.16) are essential. The results of this analysis look as follows.

The $(x^m,\theta^\alpha,L_\beta)$ cycle yields
in the zeroth order in $\theta$ and $L$
\begin{equation}
\frac{1}{2m}+d_1=O^1(T), \qquad
\frac{1}{2m}+\frac{c_1}{4im}+a_2=O^1(T).
\end{equation}
Analogous results for the $(x^m,\theta^\alpha,\theta^\beta)$- and
$(x^m,L_\alpha,L_\beta)$-cycles are
\begin{equation}
\frac{1}{8im^2}-\frac{a_1}{4im}+b_2=O^1(T), \qquad c_2=O^1(T).
\end{equation}
Evaluating the cycle $(x^m,x^n,L_\alpha)$ one can get that the terms
linear in $\theta$ and $L$ vanish if the equations
$$
c_1a_2+\frac{1}{2\sqrt{-\pi^2}}c_1+g_1=O^1(T), \qquad
\frac{1}{2\sqrt{-\pi^2}}c_1+c_1d_1=O^1(T);
\eqno{(4.19a)}$$
$$
-d_1d_2+\frac{1}{2\sqrt{-\pi^2}}(d_1-d_2)=O^1(T), \qquad
c_1b_2-d_1d_2+\frac{1}{2\sqrt{-\pi^2}}(d_1-d_2)+g_3=O^1(T)
\eqno{(4.19b)}$$
were fulfilled. Comparing now the first equations in Eqs. (4.17) and
(4.19b) one concludes that
$$
\frac{1}{m\sqrt{-\pi^2}}=O^1(T).
\eqno{(4.20)}$$
This is a contradictory equation. Thus, for the $D=9$ massive
superparticle a possibility to continue the Jacobi identity off the
constraint surface proved to be in a conflict with the manifest
Poincar\'e covariance. The following remarks are relevant here. First,
if one considers the Jacobi identity in a weak sense the contradictions
do not appear (for example, Eq. (4.19b) is a coefficient at $\chi^-$
in the $(x,x,L)$-cycle) what reproduces the result of Sec. 3.2.
Secondly, it is straightforward to check that the requirement of the
Jacobi identity in the case when the coefficients at the constraints
are supposed to be $x^m$-dependent leads to the singular fundamental
phase variable brackets.

\section{Conclusion}

In the present paper we have constructed a consistent covariant
Dirac-type bracket for the Brink--Schwarz superparticle in
arbitrary dimension. This was achieved by enlarging the original phase
space and introducing into the consideration a pair of strong
projectors (existing for described case in the extended space only)
splitting the original fermionic constraints into (infinitely)
reducible first and second class parts. The proposed bracket was
shown to possess all the rank and algebraic properties of a standard
Dirac bracket when restricted to the second class constraints surface,
what is sufficient for conventional canonical quantization of the theory.
A covariant (redundant) gauge fixing and a consistent elimination of
the second class constraints present no special problems within
the framework of the developed scheme.

A possibility to quantize the superparticle on the basis of the ``unified
constraint dynamics'' by Batalin and Tyutin was examined. As was shown,
although the classical boundary conditions for the quantization procedure
can be constructed, the covariant quantum realization of the quantities
is problematic. The latter circumstance turned out to be related to the
impossibility to continue off-shell the Jacobi identity for the constructed
bracket along the Poincar\'e covariant lines.

In this paper we have realized the general procedure presented in Sec.~2
for the superparticle models. We hope, however, that the approach
can be extended as well to the superstring and superbrane models which
possess similar problems.

\section*{Acknowledgment}
This work was supported in part by ISF Grant No M2I300 and European
Community Grant No INTAS-93-2058. The work of (A.V.G.) has been made
possible by a fellowship of Tomalla Foundation (under the research program
of ICFPM) and ISSEP Grant No A837-F.

\end{document}